\documentclass[twocolumn,showpacs,preprintnumbers,amsmath,amssymb,prb]{revtex4-1}

\usepackage[dvips,colorlinks=true,bookmarks=false,citecolor=blue,urlcolor=blue]{hyperref} 

\usepackage{amsmath,amssymb}
\usepackage{graphicx}
\usepackage{verbatim} 
\usepackage{color}

\usepackage[english]{babel}
\usepackage{bm}
\usepackage{natbib}
\usepackage{graphicx}

\def\dfrac{\displaystyle\frac}  

\renewcommand {\phi}{\varphi}

\newcommand{\eps}{\varepsilon}

\newcommand{\ve}{\varepsilon}

\newcommand{\prp}{\scriptscriptstyle{\perp}}
\newcommand{\vrt}{\scriptscriptstyle{\Vert}}

\newcommand{\nix}[1]{}

\begin{document}
\title{Dissipative plasmon solitons in graphene nanodisk arrays}

\author{Daria A. Smirnova$^1$}
\author{Roman E. Noskov$^{2,3}$}
\author{Lev A. Smirnov$^{4,5}$}
\author{Yuri S. Kivshar$^{1,2}$}

\affiliation{
$^1$Nonlinear Physics Center, Research School of Physics and Engineering,
Australian National University, Canberra ACT 2601, Australia \\
$^2$ITMO University, St.~Petersburg 197101, Russia\\
$^3$Max Planck Institute for the Science of Light, Erlangen D-91058, Germany\\
$^4$Institute of Applied Physics, Russian Academy of Sciences, Nizhny Novgorod 603950, Russia\\
$^5$Lobachevsky State University, Nizhny Novgorod 603950, Russia}

\pacs{78.67.Wj, 42.65.Tg, 42.79.Gn}


\begin{abstract}
We study nonlinear modes in one-dimensional arrays of doped graphene nanodisks with Kerr-type nonlinear response in the presence of an external electric field. We present the theoretical model describing the evolution of the disks' polarizations, taking into account intrinsic graphene losses and dipole-dipole coupling between the graphene nanodisks. We reveal that this nonlinear system can support discrete dissipative scalar solitons of both longitudinal and transverse polarizations, as well as vector solitons composed of two mutually coupled polarization components. We demonstrate the formation of stable resting and moving localized modes under controlling guidance of the external driving field.
\end{abstract}

\maketitle

\section{Introduction}

The study of plasmonic effects in graphene structures has attracted a special interest from the nanoplasmonics research community due to novel functionalities delivered by such systems, including a strong confinement by a graphene layer and tunability of graphene properties through doping or electrostatic gating~\cite{JablanPRB,Abajo176,Engheta_sci_2011,RevGrigorenko,RevBao,JablanReview,RevLuo,Abajo_Review_ACSPhot}.
Recent experiments provided the evidence for the existence of {\em graphene plasmons} revealed by means of the scattering near-field microscopy and the nanoimaging methods~\cite{Koppens_exp,Basovexp}. Being guided by a graphene monolayer, p-polarized plasmons are extremely short-wavelength, and their excitation is rather challenging. In order to decrease the plasmon wavenumbers, multilayer graphene structures can be employed~\cite{DissipSoliton_LPR, MultiWg}. Alternatively, to realize coupling of graphene plasmons with light, the in-plane momentum matching can be attained in the graphene structures with a broken translational invariance, such as graphene patterned periodically in arrays of nanoribbons~\cite{Ju_2011, bludov_primer_2013, Nikitin_2012, Nikitin_2013} or disks~\cite{Abajo182PRL, Yan_2012, AbajoACSNano}. Being regarded as direct analogs to metal nanoparticles, finite-extent nanoflakes are created by nanostructuring of graphene in the form of disks, rings, and triangles, and they can sustain localized surface plasmons. Importantly, a tight confinement of graphene plasmons results in the field enhancement indispensable for the observation  of strong nonlinear effects. In this respect, nonlinear response of graphene structures and plasmonic phenomena with graphene still remain largely unexplored.

In this paper, we study nonlinear effects in periodic arrays of single-layer graphene nanodisks excited by an external field. We assume that the nanodisks possess a nonlinear response due to the graphene nonlinearity, and demonstrate that this system can support different classes of localized modes comprising several coupled nanodisks characterized by the local field enhancement, the so-called \textit{discrete dissipative plasmon solitons}, as shown schematically in Fig.~\ref{fig:Fig1}. We derive the nonlinear equations describing the evolution of the disks' polarization components, taking into account graphene nonlinear response, intrinsic graphene losses, and a full dipole-dipole coupling between the graphene nanodisks. We reveal that this nonlinear system can support both scalar and vector discrete dissipative solitons and, depending on the inclination of the incident wave, these nonlinear modes can move gradually along the chain.
We believe that our results may be useful for initiating the experimental studies of nonlinear effects in the photonic systems with nanostructured graphene.


\section{Model}

\begin{figure}[t]
\centering\includegraphics[width=1\linewidth] {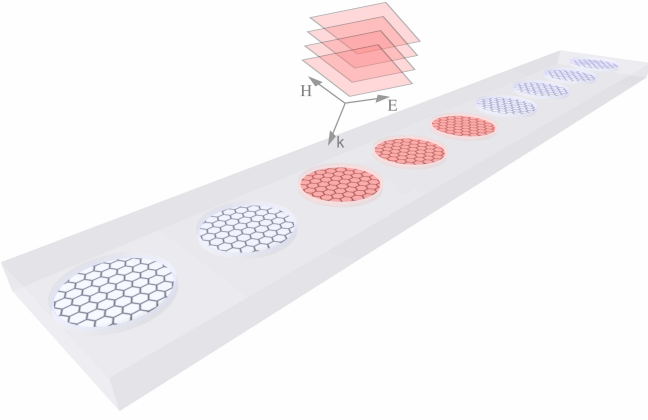} 
\caption{(Color online) Schematic view of a discrete plasmon soliton excited by an external plane wave in a chain of graphene nanodisks. Red and white colors depict high and low values of the local electric field.
}
\label{fig:Fig1}
\end{figure}

We consider an one-dimensional chain of identical graphene circular nanodisks driven by an external plane wave, as shown in  Fig.~\ref{fig:Fig1}. We assume that the radius of a single disk, $a$, varies from $15$~nm to $100$~nm, the array period $d$ satisfies the condition $d\ge3a$, and the wavelength of the driving field is much larger than a single disk, so that we can neglect boundary, nonlocal, and quantum finite-size effects~\cite{Abajo183,JablanReview}, and treat disks as point dipoles \cite{Abajo182PRL, Abajo176, Abajo_Rib_Wg, AbajoACSNano}. We also employ the linear surface conductivity as that of a homogeneous graphene sheet, which, at the relatively low photon energies, $\hbar \omega \leq \mathcal{E}_{F}\:$, can be written in terms of the Drude model as follows~\cite{Abajo194, Mikh_Nonlin, falk},
\begin{equation}
\sigma^\text{L} (\omega) = - \displaystyle{\frac{ie^2}{\pi \hbar^2}\frac{\mathcal{E}_{F}}{\left(\omega - i\tau_{\text{intra}}^{-1}\right)}}\:,
\end{equation}
where $e$ is the elementary charge, $\mathcal{E}_{F} = \hbar V_{F} \sqrt{\pi n}$ is the Fermi energy, $n$ is the doping electron density, $V_{F}\approx c/300$ is the Fermi velocity, and $\tau_{\text{intra}}$  is a relaxation time (we assume $\exp (i \omega t)$ time dependence). Hereinafter, for doped graphene we account for intraband transitions only and disregard both interband transitions and temperature effects, implying $k_{B}T \ll \mathcal{E}_{F}$, where $k_{B}$ is the Boltzmann constant and $T$ is the absolute temperature.

Under accepted approximations, the linear response of graphene nanodisks can be characterized via the disk polarizability, written as follows~\cite{AbajoACSNano},
\begin{equation}\label{eq:alpha}
\alpha^{\text{L}} (\omega) = D^3 \displaystyle{A \left( \frac{L}{\eps_h} + \frac{i \omega D}{\sigma^\text{L} (\omega)} \right)^{-1} }\:,
\end{equation}
where $A= 0.65$, $L = 12.5$, $D=2a$, $\eps_h={(\eps_1 + \eps_2)}/{2}$, $\eps_1$ and $\eps_2$ are the dielectric permittivities of the substrate and superstrate located below and upper a graphene nanodisk. Being size- and material-independent, the coefficients $A$ and $L$ are extracted from numerical simulations of Maxwell's equations by the boundary element method~\cite{Abajo182PRL, Abajo_PRB_BEM_2002, Hohenester_2012}, where graphene is modeled as a thin layer of the thickness $h = 0.5$~nm being described by volume dielectric permittivity,
\begin{equation} 
\label{eq:epsilon_gr}
\eps_{\text{gr}}^{\textnormal{L}} (\omega) =  1 - \dfrac{i 4\pi \sigma^{\textnormal{L}}(\omega) } {\omega h}\:.
\end{equation}
%
Equation~\eqref{eq:alpha} results in the following expression for the eigenfrequency of the dipole plasmon~\cite{Abajo176,AbajoACSNano}
\begin{equation}
\hbar \omega_{0}  \approx e \left( \frac{2L}{\eps_1 + \eps_2} \frac{\mathcal{E}_{F}}{\pi D} \right)^{1/2}\:.
\end{equation}
%
We notice that $\omega_{0}$ can be tuned by doping ($\mathcal{E}_{F}$-shift) or shape-cut, which may assist matching waves of different polarizations in circuits based on patterned graphene.

Within the dipole approximation, the local electric field in disks is supposed to be homogeneous. To identify it, we model disks as spheroids of the permittivity~\eqref{eq:epsilon_gr}. By comparison of Eq.~\eqref{eq:alpha} and the polarizability of an oblate spheroid~\cite{LL_EDSS}, one can conclude that the ratio between semi-minor axis $\bar c$ of an equivalent ellipsoid and the thickness $h$ should be about $\bar c/h \approx  0.627$. 
To account for the influence of graphene nonlinearity on the disks' polarizations, we 
define the nonlinear dielectric permittivity as
$\ve_{\textnormal{gr}}^{\textnormal{NL}}(\omega) =\ve_{\textnormal{gr}}^{\textnormal{L}}(\omega)  +\chi^{(3)}_{\text{gr}}(\omega) |{\bf E}^{\text{in}}_n|^2$,  where ${\bf E}^{\text{in}}_n$ is the local field in $n$-th disk, and cubic volume susceptibility,
\begin{equation} \label{eq:chi}
\chi^{(3)}_{\text{gr}} (\omega) = - i\displaystyle{\frac{  4 \pi  \sigma^{\text{NL}} (\omega)} {\omega h}} \:,
\end{equation}
is expressed through the nonlinear self-action correction to the graphene conductivity, in the local quasi-classical approximation given by~\cite{Mikh_Nonlin, Mikh_Ziegler_Nonlin, Glazov2013, Peres2014}
\begin{equation}
\label{eq:nonl_cond_gr}
\sigma^{\text{NL}} (\omega) =  i \displaystyle{\frac{9}{8} \frac{e^4}{\pi \hbar^2} \left(\frac{ V_{F}^2 } { \mathcal{E}_{F} \omega^3}\right)}\:. 
\end{equation}

Next, we study the chain of graphene nanodisks driven by an optical field with the frequency close to he frequency $\omega_0$, and analyze the dynamical response of the disks' polarizations, $p_n^{\perp,||}$. By employing the dispersion relation method~\cite{PhysRevLett.108.093901, Noskov_OE_2012, Noskov_SciRep_2012, Noskov_OL_2013}, we derive the following system of coupled nonlinear equations for the slowly varying amplitudes of the disk dipole moments,
\begin{equation}
\begin{aligned}
-i\frac{d P_n^{\prp}}{d\tau}+\left(-i\gamma+\Omega+|{\bf P}_n|^2 \right) P_n^{\prp}+ \sum_{m\neq n} G_{n,m}^{\prp} P_m^{\prp} &= E_n^{\prp}, \\
-i\frac{d P_n^{\vrt}}{d\tau}+\left(-i\gamma+\Omega+|{\bf P}_n|^2 \right) P_n^{\vrt}+ \sum_{m\neq n} G_{n,m}^{\vrt} P_m^{\vrt} &= E_n^{\vrt}, \label{eq:dynamicEQ}
\end{aligned}
\end{equation}
where
\begin{equation*}
\begin{aligned}
G_{n,m}^{\prp}  & = \frac{\eta}{2} \left( (k_0 d)^2 - \frac{i k_0 d}{|n-m|}- \frac{1}{|n-m|^2} \right) \frac{e^{-i k_0 d|n-m|}}{|n-m|}\;,\\
G_{n,m}^{\vrt}  & = \eta \left(\frac{i k_0 d}{|n-m|} + \frac{1}{|n-m|^2} \right) \frac{e^{-i k_0 d|n-m|}}{|n-m|}\:,
\end{aligned}
\end{equation*}
while
\begin{equation*}
P_n^{\prp,\vrt}=p_n^{\prp,\vrt}\frac{{3}\sqrt{\chi^{(3)}_{\text{gr}} (\omega_{0})}n_{x}}{\sqrt{2\left(1-\varepsilon_h+\varepsilon_h/{n_x}\right)}\varepsilon_h { a^2 \bar c }}\:,
\end{equation*}
and
\begin{equation*}
E_n^{\prp,\vrt} = - \frac{\varepsilon_h \sqrt{\chi^{(3)}_{\text{gr}} (\omega_{0})} E^{(ex)\prp,\vrt}_n }{{n_x}\sqrt{8\left(1-\varepsilon_h+\varepsilon_h/{n_x}\right)^3}}\
\end{equation*}
are the normalized slowly varying envelopes of the disk dipole moments and the external field, indexes '$\perp$' and '$||$' stand for transversal and longitudinal components with respect to the chain axis, $|{\bf P}_n|^2=|P_n^{\prp}|^2+|P_n^{\vrt}|^2$, $\quad \tau = \omega_0 t$, $\Omega=(\omega-\omega_0)/\omega_0$, $k_0=\omega_0\sqrt{\varepsilon_h}/c$, $c$ is the speed of light, $n_x = \pi \bar c/ 4 a$ is the depolarization factor of the ellipsoid, 
\begin{equation*}
\gamma=\frac{\nu}{2\omega_0}+\frac{\varepsilon_h }{1-\varepsilon_h+\varepsilon_h/{n_x}}\left(\frac{ k_0^3 a^2\bar c}{9{n_{x}}^{2}}\right)
\end{equation*}
describes both thermal and radiation energy losses,
\begin{equation*}
\eta=\frac{\varepsilon_h}{1-\varepsilon_h+\varepsilon_h/{n_x}}\left(\frac{ a^2\bar c}{3{n_{x}}^{2}d^3}\right)\;.
\end{equation*}
Importantly, this model involves all disk interactions through the full dipole fields, and it
can be applied to both finite and infinite chains. It should be noted that the disk polarizability across the flake plane is supposed to be zero owing to atomic-scale thickness of  graphene.

It was shown that a similar model descrbing arrays of metal nanoparticles exhibits interesting nonlinear dynamics for one- and two-dimensional arrays, including the generation of kinks, oscillons, and dissipative solitons~\cite{PhysRevLett.108.093901, Noskov_OE_2012, Noskov_SciRep_2012, Noskov_OL_2013}. Here, we focus on one-dimensional \textit{bright} localized soliton solutions similar of those are known for various discrete dissipative systems~\cite{Akhmediev_DS, Ackermann, Rosanov}.

\begin{figure}[b!]
\centering\includegraphics[width=1\linewidth]{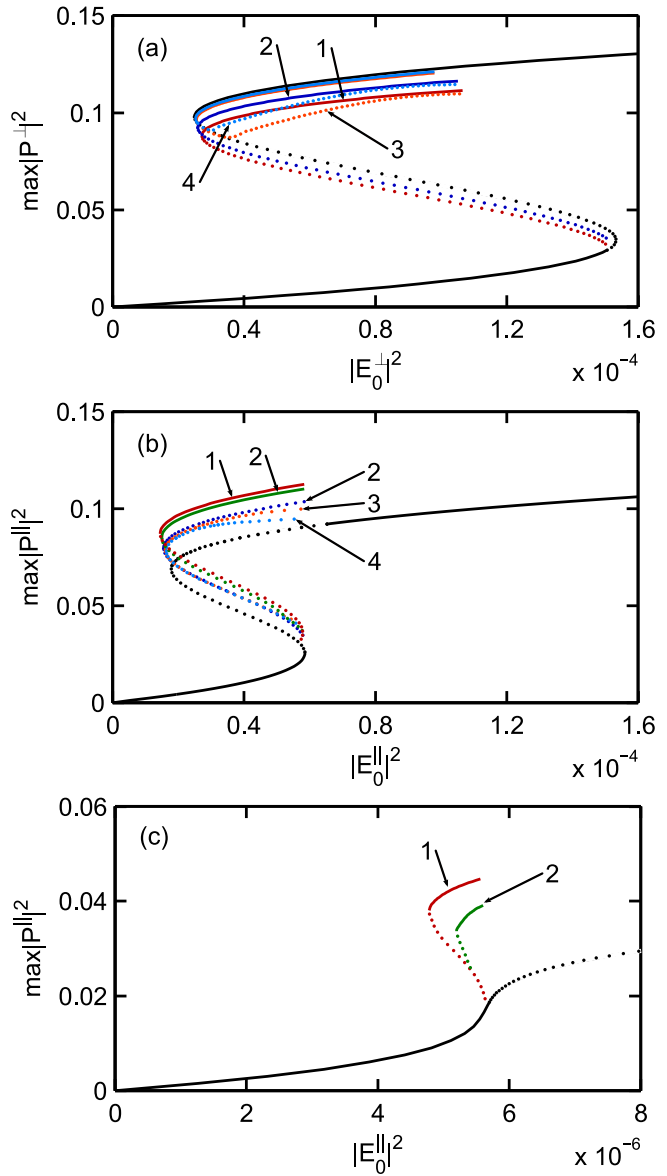} 
\caption{(Color online) Homogeneous stationary solution (black line) and soliton families. 
Black dotted indicates modulationally unstable part of the dependence. Solid and dotted color curves correspond to stable and unstable branches of solitons with different number of peaks marked by digits: (a) transversely polarized, $\Omega = -0.09$; longitudinally polarized solitons, (b) $\Omega = -0.09$, (c) $\Omega = -0.045$ (green line corresponds to the bound state). 
} 
\label{fig:Fig2}
\end{figure}

\section{Soliton families}

By varying the pump configuration, we can decouple the nonlinear equations~\eqref{eq:dynamicEQ} and analyze {\em scalar solitons} in each of the polarization components separately. However, in a general case, the polarization components 
remain coupled, and we should study the case of two-component, or {\em vector solitons}. 

In our calculations throughout this paper, we employ the following set of parameters: $a=30$ nm, $\eps_h = 2.1$, $\mathcal{E}_{F} = 0.6$ eV, $\tau_{\text{intra}} = 0.127$ ps, $d = 3.8a$, and $\hbar \omega_0 \approx 0.165$ eV (which corresponds to the wavelength $\approx 7.5$ $\mu$m). However, we notice that, within
our model, the results remain valid for a broad range of parameters, which can be adjusted for controlling the effect.

\subsection{Scalar solitons}

First, we excite the chain by a homogeneous electric field with two polarizations: (i) ${\bf E}_n=(E_n^{\vrt}, 0)$ and (ii)  ${\bf E}_n=(0, E_n^\perp)$. Assuming the driving radiation, e.g. normally incident pump plane wave, has the in-plane electric field component either across or along the chain axis, we solve the decoupled equations of the system Eqs.~\eqref{eq:dynamicEQ}. 

Since dissipative solitons are supposed to nest on a stable background, we begin with the analysis of a steady homogeneous state and inspecting its modulational stability. For an infinite chain, following~\cite{PhysRevLett.108.093901,Noskov_SciRep_2012}, we find analytically the homogeneous stationary solutions of Eqs.~(\ref{eq:dynamicEQ}) which are characterized by bistability at $\Omega<-0.047$ and $\Omega<-0.018$ for the transversal and longitudinal excitations, respectively, as shown in Figs.~\ref{fig:Fig2}(a,b). For finite chains, conclusions drawn from the analytical considerations have to be verified numerically since the edges may produce additional boundary instabilities. However, typically discrete solitons exist inside or nearby a bistability domain. Therefore, we will focus on these regions to identify soliton families.

In practice, dissipative solitons can be formed, for instance, when the chain is subject to additional narrow beam pulses.
Another way for formation of solitons is the collision of switching waves (kinks)~\cite{Noskov_OE_2012}, step-like distributions which connect quasi-homogeneous levels corresponding to the top and low branches of a bistable curve. In this way, discrete solitons are frequently interpreted as two tightly bound kinks with the opposite polarities.

\begin{figure}[t]
\centering\includegraphics[width=1\linewidth]{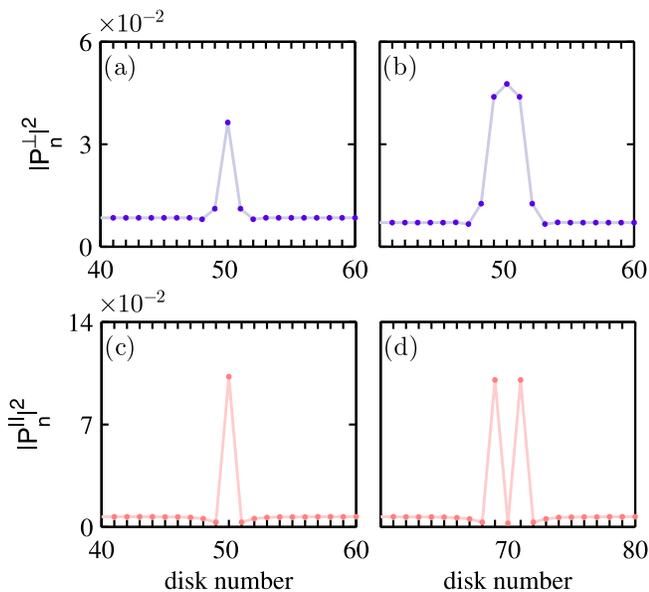}
\caption{(Color online) Soliton profiles in the case of homogeneous excitation:
(a) one-peak transverse soliton at $\Omega = -0.04$ , $|E_0^{\perp}|^2=1.58\times10^{-5}$,
(b) three-peak transverse soliton at $\Omega = -0.04$, $|E_0^{\perp}|^2=1.4\times10^{-5}$,
(c) one-hump longitudinal soliton coexisting with a bound state (d) containing two peaks at $\Omega = -0.09$, $|E_0^{\vrt}|^2=3\times10^{-5}$, sitting on the background of a homogeneous steady state solution.
}
\label{fig:Fig3}
\end{figure}

Applying the standard Newton iteration scheme for a finite chain of 101 disks, we find families of bright solitons, characterized by a snaking bifurcation behavior~\cite{Ackermann, Lederer_OL_2004, Egorov_OE_2007,Rojas}, and simultaneously determine their stability, as shown in Fig.~\ref{fig:Fig2}. Remarkably, longitudinal solitons also appear outside the bistability area, where a homogeneous steady state solution is a single-valued function of the pump, provided that the character of the bifurcation is subcritical, particularly, in the certain range of frequencies, $ -0.047 < \Omega < -0.04$, for the parameters of Fig.~\ref{fig:Fig2}. Examples of the soliton profiles are depicted in Fig.~\ref{fig:Fig3}. 

Within the homogeneous excitation, solitons always stand at rest regardless their width
because the effective periodic potential created by the chain requires a finite value of the applied
external force to start soliton's motion~\cite{Rosanov,Egorov_OE_2007}. In order to study soliton mobility, we excite the chain by the tilted light incidence: ${\bf E}_n=(E_0^{\vrt} \exp(-i Q d n), 0)$ and ${\bf E}_n=(0, E_0^\perp \exp(-i Q d n))$, where $Q$ is the longitudinal wavenumber.

\begin{figure}[t]
\centering\includegraphics[width=1\linewidth]{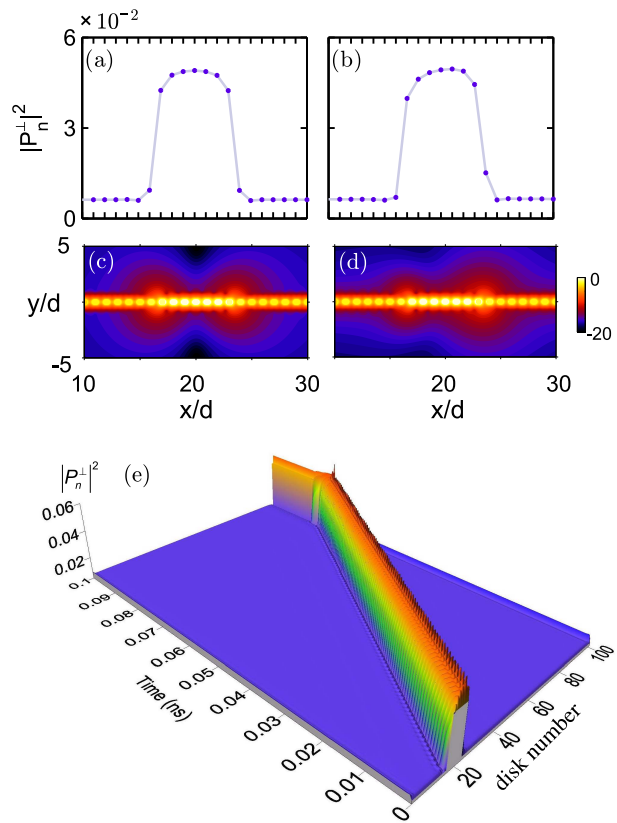}
\caption{(Color online)
Soliton profiles at $|E_0^{\perp}|^2=1.28\times10^{-5}$, $\Omega=-0.04$, (a) $Qd=0$ symmetric resting,
(b) $Qd= - 0.05$ asymmetric resting; (c,d) respective top views of the intensity distribution, $\ln \left(|E|^2/|E|_{\text{max}}^2\right)$, in the plane of the chain; (e) Spatiotemporal dynamics of a drifting soliton at $Qd=-0.2$. 
}
\label{fig:Fig4}
\end{figure}

In contrast to cavity solitons in a model with the nearest-neighbor purely real coupling and focusing
nonlinearity~\cite{Egorov_OE_2007}, the longitudinally polarized one-peak solitons remain trapped at any value of 
$Q$. This is associated with the imaginary part of the dipole-dipole interaction. However, wide enough transverse solitons are susceptible to a propelling force. Example of multi-peaked moving solitons is presented in Fig.~\ref{fig:Fig4}.
At $Qd\neq 0$, the soliton looses its symmetry [see Figs.~\ref{fig:Fig4}(b,d)] and, in the presence of an in-plane 
momentum exceeding some critical value, the soliton starts moving along the chain towards the edge where it gets trapped
as shown in Fig.~\ref{fig:Fig4}(e).

\subsection{Vector solitons}

\begin{figure}[h!]
\centering\includegraphics[width=1\linewidth]{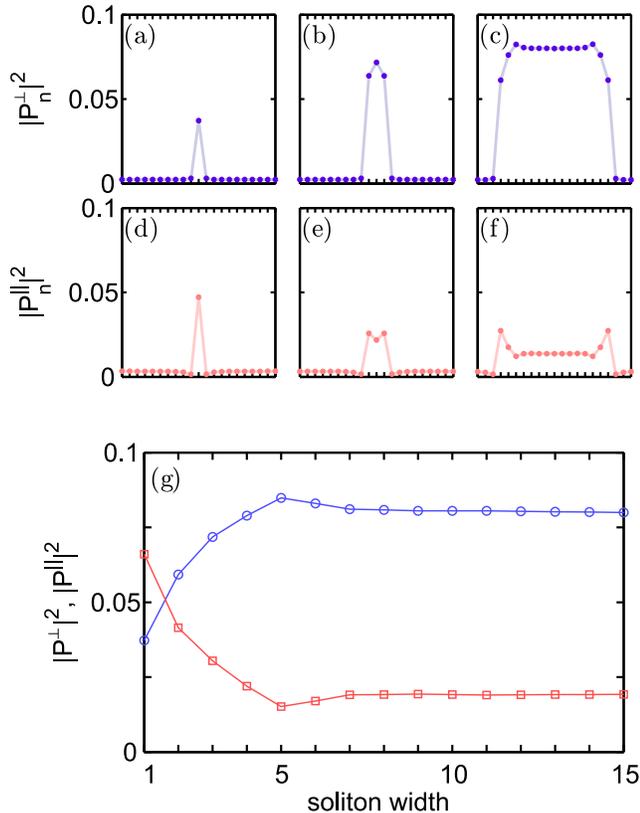}
\caption{(Color online) Profiles of two-component vector solitons at $|E_0|^2=4\times10^{-5}$, $\Omega=-0.09$, $\theta=0.25 \pi$ localized on (a,d) one and (b,e) three nanodisks. The case (c,f) shows an example of a broad vector soliton.
(g) Dependence of the amplitude in the central excited disk of the vector soliton on the number of excited disks.
Circles and squares correspond to the transverse and longitudinal components, respectively. 
}
\label{fig:Fig5}
\end{figure}

Remarkably, coupled equations~\eqref{eq:dynamicEQ} also support two-component vector solitons with a mixed polarization when the excited field contains both nonzero components, $E_0^{\prp,\vrt} \neq 0$, $ {\textbf E}_0 =  (E_0^{\prp}, E_0^{\vrt}) = (E_0 \sin \theta, E_0 \cos \theta) $, see examples in Fig.~\ref{fig:Fig5}. In the case of vector solitons, horned longitudinal solitons corresponding to the dotted unstable branches in Fig.~\ref{fig:Fig2}(b), become stabilized. Figure~\ref{fig:Fig5}(g) illustrates a variation of the amplitudes of both components with the growing soliton width.

\section{Concluding remarks}

In our analysis presented above, we have operated with dimensionless variables. To estimate the feasibility of the predicted phenomena, we should recover physical values and realistic parameters.
In particular, Eqs.~\eqref{eq:chi},~\eqref{eq:nonl_cond_gr} provide the estimate of $\chi^{(3)}_{\text{gr}} \sim 10^{-7} - 10^{-6}$~esu for typical parameters, and the dimensionless intensity of the external field, $|E_0^{\prp, \vrt}|^2 \sim 10^{-4}$, corresponds to the physical intensity of $\sim 10 $~kW/cm$^2$. Even accounting for the local field enhancement inside the graphene nanodisks, the characteristic time scales at which the solitons are formed are estimated to be less than the pulse duration that may cause a graphene damage at the given intensities~\cite{damage_Krauss_APL_2009, damage_Currie_APL_2011, damage_Roberts_APL_2011}. Thus, there are expectations for the experimental observation of the predicted nonlinear effect.

In summary, we have studied nonlinear dynamics in nonlinear arrays of graphene nanodisks in the presence of a pumped external field. We have derived nonlinear equations describing the evolution of the nanodisks' polarization, taking into account losses in graphene and a dipole-dipole coupling between the nanodisks. We have demonstrated the existence of families of discrete dissipative solitons and also revealed that such solitons can propagate stably along the chain provided they are excited by the tilted field. We have predicted a new class of discrete vector solitons composed of two mutually coupled polarization components. Our findings may pave a way to the soliton-based routing in optoelectronic circuits based on nanostructured graphene.

\acknowledgements{This work was supported by the Ministry of Education and Science of Russia
(project 14.124.13.6087-MK), Government of the Russian Federation (grants no. 074-U01, 14-12-00811),
and the Australian National University.}


\begin{thebibliography}{100}

\bibitem{JablanPRB} M. Jablan, H. Buljan, and M. Soljacic, Phys. Rev. B \textbf{80}, 245435 (2009).

\bibitem{Abajo176} F. H. L. Koppens, D. E. Chang, and F. J. Garcia de Abajo, Nano Lett. \textbf{11}, 3370 (2011).

\bibitem{Engheta_sci_2011} A. Vakil and N. Engheta, Science \textbf{332}, 1291 (2011).

\bibitem{RevGrigorenko} A.~N. Grigorenko, M. Polini, and K. S. Novoselov, Nature Photon. \textbf{6}, 749 (2012).

\bibitem{RevBao} Q. Q. Bao and K. P. Loh, ACS Nano \textbf{6}, 3677 (2012).

\bibitem{JablanReview} M.~Jablan, M.~Soljacic, and H.~Buljan, Proc. IEEE \textbf{101}, 1689 (2013).

\bibitem{RevLuo}
X. Luo, T. Qiu, W. Lu, and Z. Ni, Mat. Sci. Eng. R \textbf{74}, 351 (2013).

\bibitem{Abajo_Review_ACSPhot} F. J. Garcia de Abajo, ACS Photonics \textbf{1}, 135 (2014).

\bibitem{Koppens_exp}
J.~Chen, M.~Badioli, P.~Alonso-Gonzales, S.~Thongrattanasiri, F.~Huth, J.~Osmond, M.~Spasenovic, A.~Centeno, A.~Pesquera, P.~Godignon, A.\,Z. Elorza, N.~Camara, F.\,J.\,G. de Abajo, R.~Hillenbrand, and F.\,H.\,L. Koppens, Nature \textbf{487}, 77 (2012).

\bibitem{Basovexp}
Z. Fei, A.\,S. Rodin, G.\,O. Andreev, W. Bao, A.\,S. McLeod, M. Wagner, L.\,M. Zhang, Z.~Zhao, G.~Dominguez, M.~Thiemens, M.\,M. Fogler, A.\,H. Castro-Neto, C.\,N. Lau, F.~Keilmann, and D.\,N. Basov, Nature \textbf{487}, 82 (2012).


\bibitem{DissipSoliton_LPR} D.\,A. Smirnova, I.\,V. Shadrivov, A.\,I. Smirnov, and Y.\,S. Kivshar, Laser Photonics Rev. \textbf{8}, 291 (2014).

\bibitem{MultiWg}  D.\,A. Smirnova, I.\,V. Iorsh, I.\,V. Shadrivov, and Y.\,S. Kivshar, JETP Lett. \textbf{99}, 456 (2014).

\bibitem{Ju_2011} L. Ju, B. Geng, J. Horng, C. Girit, M. Martin, Z. Hao, H. A. Bechtel, X. Liang, A. Zettl,
Y. R. Shen, and F. Wang,  Nature Nanotechnol. \textbf{6}, 630 (2011).

\bibitem{bludov_primer_2013} Yu. V. Bludov, A. Ferreira, N. M. R. Peres, and M. I. Vasilevskiy, Int. J. Mod. Phys. B \textbf{27}, 1341001 (2013).

\bibitem{Nikitin_2012} A. Yu. Nikitin, F. Guinea, F. J. Garcia-Vidal, and L. Martin-Moreno, Phys. Rev. B \textbf{85}, 081405 (2012).

\bibitem{Nikitin_2013} T. M. Slipchenko, M. L. Nesterov, L. Martin-Moreno and A. Yu. Nikitin, J. Opt. \textbf{15} (2013).

\bibitem{Abajo182PRL} S. Thongrattanasiri, F. H. L. Koppens, and F. J. Garcia de Abajo, Phys. Rev. Lett. \textbf{108}, 047401 (2012).

\bibitem{Yan_2012} H. Yan, X. Li, B. Chandra, G. Tulevski, Y. Wu, M. Freitag, W. Zhu, P. Avouris, and F. Xia, Nature Nanotechnol. \textbf{7}, 330 (2012).

\bibitem{AbajoACSNano} Z. Fang, S. Thongrattanasiri, A. Schlather, Z. Liu, L. Ma, Y. Wang, P. M. Ajayan, P. Nordlander, N. J. Halas, and F. J. Garcia de Abajo, ACS Nano \textbf{7}, 2388 (2013).

\bibitem{Abajo183} S. Thongrattanasiri, A. Manjavacas, and F. J. Garcia de Abajo, ACS Nano \textbf{6}, 1766 (2012).	

\bibitem{Abajo_Rib_Wg} J. Christensen, A. Manjavacas, S. Thongrattanasiri, F. H. L. Koppens, and F. J. Garcia de Abajo, ACS Nano \textbf{6}, 431 (2012),

\bibitem{Mikh_Nonlin} 
S.\,A. Mikhailov, Europhys. Lett. {\bf 79}, 27002 (2007).

\bibitem{falk} L.\,A. Falkovsky and A.\,A. Varlamov, Eur. Phys. J. B \textbf{56}, 281 (2007).

\bibitem{Abajo194} S. Thongrattanasiri, I. Silveiro, and F. J. G. de Abajo, Appl. Phys. Lett. \textbf{100}, 201105 (2012).

\bibitem{Mikh_Ziegler_Nonlin} 
S.\,A. Mikhailov and K.~Ziegler, J. Phys.: Condens. Mat. \textbf{20}, 384204 (2008).

\bibitem{Glazov2013} M.\,M. Glazov and S.\,D. Ganichev, Physics Reports 535, 101-138 (2014).

\bibitem{Peres2014} 
N. M. R. Peres, Yu. V. Bludov, Jaime E. Santos, Antti-Pekka Jauho, and M. I. Vasilevskiy, Phys. Rev. B \textbf{90}, 125425 (2014).
	
\bibitem{Abajo_PRB_BEM_2002} F. J. Garcia de Abajo and A. Howie, Phys. Rev. B \textbf{65}, 115418 (2002).

\bibitem{Hohenester_2012} U. Hohenester and A. Trugler, Computer Physics Communications \textbf{183}, 370 (2012).

\bibitem{LL_EDSS} L. D. Landau and E. M. Lifshitz, {\em Electrodynamics of Continuous Media} (Oxford: Pergamon Press, 1984).

\bibitem{PhysRevLett.108.093901} R. E. Noskov, P. A. Belov, and Y. S. Kivshar, Phys. Rev. Lett. \textbf{108}, 093901 (2012).

\bibitem{Noskov_OE_2012} R. E. Noskov, P. A. Belov, and Yu. S. Kivshar, Opt. Exp. \textbf{20}, 2733 (2012).

\bibitem{Noskov_SciRep_2012} R. E. Noskov, P. A. Belov, and Yu. S. Kivshar, Sci. Rep. \textbf{2}, 873 (2012).

\bibitem{Noskov_OL_2013} R. E. Noskov, D. A. Smirnova, and Yu. S. Kivshar, Opt. Lett. \textbf{38}, 2554 (2013).

\bibitem{Akhmediev_DS} N. Akhmediev and A. Ankiewicz, {\em Dissipative Solitons} (Springer-Verlag, Heidelberg, 2008).

\bibitem{Ackermann} T. Ackemann, W. J. Firth, and G. Oppo, in: {\em Advances in Atomic, Molecular, and Optical Physics}, Vol. 57 (Elsevier, 2009), pp. 323-421.

\bibitem{Rosanov} N. N. Rosanov, {\em Spatial Hysteresis and Optical Patterns} (Springer-Verlag, Heidelberg, 2002).

\bibitem{Lederer_OL_2004} U. Peschel, O. Egorov, and F. Lederer, Opt. Lett. \textbf{29}, 1909 (2004).

\bibitem{Egorov_OE_2007} O. Egorov, F. Lederer, and Yu. S. Kivshar, Opt. Exp. \textbf{15}, 4149 (2007). 

\bibitem{Rojas} M. G. Clerc, R. G. El\'ias, and R. G. Rojas, Phil. Trans. R. Soc. A \textbf{369}, 412 (2011).

\bibitem{damage_Krauss_APL_2009} B. Krauss, T. Lohmann, D. H. Chae, M. Haluska,
K. vonKlitzing, and J. H. Smet, Phys. Rev. B \textbf{79}, 165428 (2009).

\bibitem{damage_Currie_APL_2011} M. Currie, J. D. Caldwell, F. J. Bezares, and J. Robinson, Appl. Phys. Lett. \textbf{99}, 211909 (2011).

\bibitem{damage_Roberts_APL_2011} A. Roberts, D. Cormode, C. Reynolds, T. N. Illige, B. J. Leroy, and A. Sandhu, Appl. Phys. Lett. \textbf{99}, 051912 (2011).

\end{thebibliography}

\end{document}